\documentclass[preprint,review]{elsarticle}

\usepackage{lipsum}

\usepackage{lineno}

\journal{Mathematical Biosciences}

\usepackage[utf8]{inputenc}
\usepackage{graphicx}
\usepackage{epstopdf}
\usepackage{comment}
\usepackage {subfigure}
\usepackage{subfigure}
\usepackage{amssymb,amsmath,amsthm,scalefnt}
\usepackage{mathrsfs,color}
\usepackage{geometry}
\usepackage{amsthm}
\usepackage{amsfonts}
\usepackage{mathrsfs}
\usepackage{url}

\usepackage{float}
\usepackage{hyperref}

\def\th{\theta}
\def\t{\textrm}
\let\mathcal\mathscr
\newcommand{\Rb}{\mathbb R}

\newcommand{\SUM}{\sum\limits}

\def\p{\partial}

\theoremstyle{plain}

\theoremstyle{definition}

\newtheorem* {obs}{Remark}

\begin{document}

\begin{frontmatter}
\title{Forecasting COVID-19 Chile's second outbreak by a generalized SIR model with constant time delays and a fitted positivity rate}

\author[pc,cebib]{Patricio Cumsille\corref{ca}}
\ead{pcumsille@ubiobio.cl}
\address[pc]{Department of Basic Sciences, Faculty of Sciences, University of B\'\i o-B\'\i o \\
Campus Fernando May, Av. Andr\'es Bello 720, Casilla 447, Chill\'an, Chile.}
\author[or1,or2]{\'Oscar Rojas-Díaz}
\ead{oscar.rojas.d@usach.cl, ceo@neurovision.ai}

\cortext[ca]{Corresponding author}
\address[or1]{Neurovision AI, San Eugenio 3494, Santiago, Chile.}
\address[or2]{Department of Informatics Engineering, Faculty of Engineering, University of Santiago of Chile, Santiago, Chile.}
\author[cebib]{Pablo Moisset de Espan\'es}				
\ead{pmoisset@ing.uchile.cl}

\address[cebib]{Centre for Biotechnology and Bioengineering (CeBiB) \\
University of Chile, Beaucheff 851, Santiago, Chile.}				

\begin{abstract}
The COVID-19 disease has forced countries to make a considerable collaborative effort between scientists and governments to provide indicators to suitable follow-up the pandemic's consequences. Mathematical modeling plays a crucial role in quantifying indicators describing diverse aspects of the pandemic. Consequently, this work aims to develop a clear, efficient, and reproducible methodology for parameter optimization, whose implementation is illustrated using data from three representative regions from Chile and a suitable generalized SIR model together with a fitted positivity rate. Our results reproduce the general trend of the infected's curve, distinguishing the reported and real cases. Finally, our methodology is robust, and it allows us to forecast a second outbreak of COVID-19 and the infection fatality rate of COVID-19 qualitatively according to the reported dead cases.

\section*{Highlights}
\begin{itemize}
\item We developed a clear, efficient, and reproducible methodology for parameter optimization to calibrate real data.
\item We also provided a suitable generalized SIR model, including a fitted positivity rate.
\item Our methodology reproduces the general trend of the infected' curve for both the reported and the real cases.
\item It can also forecast different scenarios of the spreading and the infection fatality rate of COVID-19 qualitatively.
\end{itemize}
\end{abstract}
\begin{keyword}
COVID-19 \sep Forecasting \sep Generalized compartmental modeling \sep Parameter identifiability methodology
\end{keyword}

\end{frontmatter}

\section{Introduction}

From the early stage of the pandemic, the Chilean government has strengthened the health system by increasing the number of intensive care unit beds and mechanical ventilation devices for critically ill patients. Currently, the government is enhancing tools for tracing capture on cases and contacts of confirmed COVID-19 patients. A collaborative effort among the main universities, the ministries of health and science, technology, knowledge, and innovation produced a web platform for data analysis. The site provides key indicators to represent the state and evolution of the pandemic. These indicators are meant to provide timely and clear information to the authorities, communication media, citizenship, and the scientific community to understand the pandemic evolution and improve decision-making.

Mathematical modeling plays a crucial role in quantifying indicators
describing diverse aspects of the pandemic. Many research groups in
Chile have developed mathematical models, and providing a comprehensive review would ask for much unnecessary effort. Therefore, we selected some that have produced periodic reports on COVID-19, which we comment on next and partly use ideas. It is worth noting that, in general, these works do not provide details on the methodology, which would be desirable to reproduce results scientifically.

The Center for Mathematical Modeling (CMM) group provided mainly simulations, not predictions. Their results aimed to estimate the maximum capacity required for intensive care unit beds and simulate different opening scenarios  (for teaching and economic activities). They developed complex modeling based on a ge\-neralization of the SEIR compartmental model, including variables for hospitalized, separating those requiring critical services, and even age structures, and containing several parameters making it difficult to deal with in the sense of forecasting; see the website. \cite{CMM}.

The COVID-19 University of Concepción (UdeC) group provided simulations and predictions. Their results aim to predict the curve of real infected for different datasets encompassing the Ñuble, Bío-Bío regions, and the entire country and simulate different scenarios with or without confinements. They developed a variant of the SEIR compartmental model, adding a fraction to account for the proportion of real infected observed by the RT-PCR tests and carrying out parameter estimation to calibrate each dataset. In our work, we used some of their ideas and compared some of our results of calibration for the Ñuble region; see the website \cite{Cabrera}.

We want to note that the research group of the previous paragraph and other manuscripts \cite{candia} support their ideas in the fact that there are unreported cases or that only a fraction of the real infected is observed. Indeed, data is measured with great uncertainty, as we explain in Section \ref{modeling}, and consequently has been reviewed and changed retrospectively. In this sense, we aim to obtain a suitable model that fits the data and forecast the general trend of the real infected' curve, despite the uncertainty in data. We believe that the way is not constructing a complex model. In that case, the parameter estimation required to make the model calibration becomes very difficult or even impossible to do in real-time or does not necessarily provide suitable results. Concerning this, we cite the work \cite{G-N} that used a complex generalization of the SEIR model proposed in \cite{Peng}, and applied it Chilean data until April 14, 2020, obtained that the peak of infected would occur between April and May, an unsuitable prediction that may be explained because data were near to its zero equilibrium, as we will discuss in Section \ref{results}.

Parameter estimation involves solving an inverse problem: given a model and measurements of some state or output variables, the parameters that characterize the system, i.e., those producing a good fit of the model with the data, need to be identified \cite{Cum-et-al-19, Cum-et-al-20}. This problem is difficult since no unique analytical or numerical solution is usually available \cite{Cum-et-al-19, Cum-et-al-20}. Even if a unique solution was available, suitable initial guesses are required by optimization solvers to compute suitable parameter estimates \cite{Cum-et-al-19, Cum-et-al-20}. From the previous discussion, the main goal of the present study was to describe a hybrid approach \cite{Cum-et-al-15}, i.e., a methodology that holistically mixes mathema\-tical modeling and experimental design, which is required, as shown in the literature, for better understanding the studied system by fitting parameters of a given model with a specific scenario and for obtaining models with predictive capability. In particular, we aim to develop a clear, efficient, and reproducible methodology for optimization parameters, whose implementation is illustrated by using data from three representative regions from Chile and a suitable generalized SIR model. The methodology relies on a numeric procedure called the Trust-Region-Reflective optimization algorithm. Simultaneously, we computed the model's predictive power using goodness-of-fit criteria and assessed parameter uncertainty through sensitivities that yield the standard errors.

\section{A generalized SIR model with constant time delays and fitted positivity rate}
\label{modeling}

The classical SIR model's main assumptions are that the mean infection and recovery rates are positive constants over time, which is not necessarily realistic for a given disease. In COVID-19, when a susceptible has just been infected does not instantaneously show symptoms. There should be a time-delay period due to the SARS-CoV-2 incubation. Similarly, the infected subjects are not recovered or dead instantaneously, but this would occur within a time-delay period. Indeed, these time-delay periods for both phenomena, incubation, and removed (by recovery or dead), are well documented and reported, e.g., in \cite{China-WHO}.

In this work, we consider the following generalization of the classical SIR model (adapted from the model from Book by Hairer et al. \cite{Hairer} p. 295; \cite{Maleewong2020}), which we will call generalized SIR model with constant time delays:
\begin{subequations}
\label{mod-base}
\begin{eqnarray}
\frac{dS}{dt}(t) &=&-\frac{\beta(t)}N S(t)I(t-\tau_1), \label{Su} \\
\frac{dI}{dt}(t) &=&\frac{\beta(t)}N S(t)I(t-\tau_1)-\gamma I(t-\tau_2), \label{I} \\
\frac{dR}{dt}(t) &=& \gamma I(t-\tau_2). \label{R}
\end{eqnarray}
\end{subequations}
The parameters of the model \eqref{mod-base} are $\beta(t)$ that corresponds to the average number of contacts per person per time, multiplied by the probability of disease transmission in exposure between a susceptible and an infectious subject (see first report at the web site \cite{CMM}), or more simply, $\beta(t)$ is the mean rate of disease transmission; $\gamma$ is the mean removal rate, which in the classical SIR model represents the average time of infection duration; $\tau_1$ is the mean incubation time of disease; $\tau_2$ is the median time from onset to clinical recovery or death caused by disease, or the duration time of disease until recovery or death.

\textcolor{black}{Parameter $\beta(t)$ depends essentially on two factors:
the disease characteristics and the contact rate within the population
\cite{refId0}. Since we cannot modify the disease characteristics,
authorities can foster or even impose incentives to each individual to
reduce his/her contact rate with other individuals in the population,
causing a decrease in $\beta(t)$. Consequently, we consider $\beta(t)$
variable in time to consider mitigation strategies dynamics of COVID-19
such as scenarios confinement and opening at different periods.}

The variables of model \eqref{mod-base} are:
\begin{enumerate}
\item $S$, the number of susceptible individuals;
\item $I$, the number of infectious individuals;
\item $R$, the number of removed (and immune), or deceased individuals.
\end{enumerate}
The initial conditions have to satisfy $S(t_0)+I(t_0)+R(t_0)=N$, where $N$ is the size of the population under study for a closed system and taking into account that $(S+I+R)'(t)=0$ for all $t >0$, and where $t_0$ is a day chosen suitably after the pandemic began, and which varies for every studied population (in Chile, the first infected subject was detected on March 02, 2020). We remark that the model \eqref{mod-base} considers the symptomatic and presymptomatic subjects since the term $I(t-\tau_1)$ at the right-hand side of the equation \eqref{Su} represents the number of infected cases that do not manifest symptoms yet since the infection is under incubation.

The solution of model \eqref{mod-base} will be defined in $[t_0,\:T[$, where $T$ is defined as the first time at which the infected number becomes zero, i.e., $I(T)=0$. From equation \eqref{I}, we have that for every $t\in[t_0,\:T[$, $I'(t)>0$ if and only if
$$\frac{\beta(t)}N S(t) I(t-\tau_1)-\gamma I(t-\tau_2)>0,$$
or equivalently
\begin{equation}
\label{Re}
R_e(t):=R^{SIR}_e(t) R_I(t)>1, \quad \t{where} \quad R^{SIR}_e(t):=\frac {\beta(t)S(t) }{\gamma N} \quad 
\t{and} \quad R_I(t):= \frac {I(t-\tau_1) }{I(t-\tau_2) }\;.
\end{equation}
In equation \eqref{Re}, $R_e(t)$ is the effective reproductive number of the generalized SIR model with constant time delays \eqref{mod-base}, $R^{SIR}_e(t)$ is the effective reproductive number in the classical SIR model, and $R_I(t)$ is the infected ratio of the latent over the removed infected.

It is worth to note that the generalized SIR model with constant time delays can generate a complex dynamics since, as we will see in Section \ref{results}, by contrast to the classical SIR model, it can simulate more than one local maximum, providing thus a way to explain a COVID-19 second-outbreak, as already observed in some European countries.

We assume that infectious subjects are detected only when RT-PCR tests are applied to them, and their results are positive. Therefore the observation of this variable is equivalent to the number of positive RT-PCR tests. In this sense, the number of asymptomatic subjects in the population is underestimated, given that they are difficult to detect. Therefore this variable is difficult to observe with accuracy, so we did not consider it in our modeling.
\textcolor{black}{\begin{obs}
The generalized SIR model with constant time delays would produce better
prediction results than the classical SEIR model since no observation of
the exposed population is available. This is because, similarly to the
asymptomatic population, the exposed are difficult to observe. Indeed,
the Chilean government database \cite{data-nat, data-reg}, apart from the symptomatic cases, counts the
asymptomatic cases, and there is no way to know how many become
symptomatic ones.
\end{obs}  }
By contrast, the symptomatic cases are detected as symptoms manifest
themselves, and the RT-PCR test confirms the infection. Therefore, this
variable is relatively well-observed, provided that the RT-PCR tests are
reliable and enough are available. Taking this into account, we assume
that the daily number of infected with symptoms reported by the ministry
of health, denoted by $I_r(t)$, is underestimated since it depends on the
availability and proper application of RT-PCR tests; see the presentation
on COVID-19 in Chile in \cite{Baeza}. Consequently, we assume that
$I_r(t)$ is a fraction of the actual number of infected cases $I(t)$,
i.e.,
\begin{equation}
\label{fr}
I_r(t) = f(t) I(t).
\end{equation}
\textcolor{black}{In equation \eqref{fr} above, $f(t)$ is the ratio
between positive RT-PCR tests number and the real infected cases in day
$t$, which accounts for the real positivity rate.} We modeled $f(t)$ in
the same way as Cabrera-Vives et al.; see the first report in
\cite{Cabrera}. That is to say, $f(t)$ is an inverted Sigmoid-type
function such that if $I(t)$ is small enough, which occurred during the
beginning of the outbreak, then an important fraction of the real
infected cases are detected ($I_r(t)\approx I(t)$). On the contrary, when
$I(t)$ is large enough, which occurred just before the quarantines were
imposed, then only a small fraction $0<a<1$ of the real infected cases
are detected ($I_r(t) \approx a I(t)$). Precisely, $f(t)$ is defined as
\begin{equation}
\label{exp-fr}
f(t) = 1 + \frac{a-1}{ 1+e^{ -k(I(t)-I_m) } }
\end{equation}
whose parameters are $a,\:k$ and $I_m$, where $a$ and $k$ represent the minimum and the decay rate of $f(t)$, respectively. On the other hand, the measure of how large/small is $I(t)$ is given by a threshold $I_m$ such that $I(t) \ll I_m$ implies $f(t) \approx 1$, and $I(t) \gg I_m$ implies $f(t)\approx a$.

\section{Material and methods}

This section describes the data and the methodology used for estimating the optimal parameters that produce the model' fit to the data.

The data corresponds to the reported Chilean government's daily official cases at the regional level in the current year, which we used to predict the COVID-19 spreading. We chose data from some representative regions of the north, south, and central zones. Specifically, we worked with the Antofagasta, the Metropolitan, and the Ñuble regions, where the Metropolitan Region is the most populous and which capital is Santiago of Chile, the country's capital. The data we used is available from the web site \cite{data-reg}; also, the countrywide level data is at  \cite{data-nat}. The data in \cite{data-reg} contains not only the number of confirmed cases but also the size of targeted populations ($N$) and the periods of quarantines per commune.

\subsection{Scenarios for the fittings and forecastings}
\label{sum}

We fitted the model \eqref{mod-base} to each dataset by taking into account different scenarios that encompass, at least, three phases of the pandemic spreading: an early stage characterized by low dissemination and levels of daily new infected cases relatively low, followed by a fast-propagation under quarantine, a measure taken by the ministry of health as a reaction to the first phase in a selective way by sectors that may encompass several communes of the region at different periods, characterized by a fast increasing of daily new infected cases and finally, a spreading slowdown stage where the quarantine is relaxed. In a general way, to account for the phases described, we modeled the mean rate of COVID-19 transmission $\beta(t)$ by a piecewise constant function as
\begin{equation}
\label{beta}
\beta(t) = \sum_{i=1}^n \beta_i \textbf{1}_{[t_{i-1},\:t_i[}(t) \qquad \t{for every } t \in [t_0,\:T[\;.
\end{equation}
In the equation \eqref{beta} above, $\beta_i$ denotes the mean rate of transmission for every scenario $i=1,\cdots,n$, $n \geq 3$ is the number of scenarios, $\textbf{1}_{[t_{i-1},t_i[}(t)$ stands for the indicator (or characteristic) function of the time interval $[t_{i-1},t_i[$ that corresponds to the $i-$th scenario, $t_0$ is the initial time, and $T$ is the maximal time of existence of model solutions, defined in Section \ref{modeling}. For the three datasets considered, the initial condition was imposed at a day $t_0$ chosen with respect to day 1 that corresponds to March 03, 2020, when it began to measure the infected in the three targeted regions.

From the definition \eqref{beta} above, we denote the model parameters as the vector 

$\th=(\beta_1,\: \beta_2, \: \cdots,\: \beta_n, \: \gamma,\: \tau_1, \: \tau_2, \: a,\: k,\: I_m)\in \Rb^{n+6}$, where $n=3$ is the minimum of scenarios to calibrate the model \eqref{mod-base} to the data. In our computations, for the Antofagasta Region dataset, we split the second phase into two to capture the real behavior of COVID-19 expansion more accurately due to two differentiated periods of quarantine in that region, taking $n=4$ in that case.

Finally, we simulated a possible change of scenario, which was evaluated by computing the relative errors predicting data after the final calibration time. To do that, we solved the model \eqref{mod-base} by taking a mean rate of transmission $\beta(t)$, as in \eqref{beta}, adding one more scenario to the $n$ computed from the calibration, and letting vary $\beta_{n+1}$ to verify the value that forecasts the best possible the data after the calibration. To be precise, we solved the model \eqref{mod-base} by redefining $\beta(t)$ as
\begin{equation}
\label{beta-sim}
\beta(t) = \sum_{i=1}^n \hat \beta_i \textbf{1}_{[t_{i-1},\:t_i[}(t) + \beta_{n+1}  \textbf{1}_{[t_{n},\:t_{n+1}[}(t)\qquad \t{for every } t \in [t_0,\:T[\;,
\end{equation}
where $(\hat \beta_1,\: \hat \beta_2,\: \cdots, \: \hat \beta_n) \in \Rb^n$ denotes the parameters estimated for the fitting of every dataset (computed together with $(\hat \gamma, \:\hat \tau_1,\: \hat \tau_2, \: \hat a, \: \hat k,\: \hat I_m)$), the time $t_n$ represents a new change of scenario, $t_{n+1}$ extends until the last predicted datum, and $\beta_{n+1}$ describes the value of the mean rate of transmission after the time $t_n$ ($\beta_{n+1}=\hat \beta_n$ means no change of scenario).

\subsection{Identifiability analysis of model' parameters}
\label{sec:2}

In this section, we explain in detail the methodology to conduct the identifiability analysis of model' parameters, i.e., finding the parameters of the generalized SIR model with constant time delays \eqref{mod-base} that best fit to the every dataset.

\subsubsection{Direct and inverse problem}

Given a parameters vector $\theta \in \Rb^{n+6}$, the \textit{direct problem} consists of finding the (unique) solution $(S, I, R)(t,\th)$ of model \eqref{mod-base} with initial condition imposed at a suitable initial time, after the pandemic began. The solution to the direct problem is required to solve the inverse problem. The present study aimed to forecast and simulate different scenarios of COVID-19 expansion in Chile, for which we developed a clear, efficient, and reproducible methodology to solve the parameter identification problem. This procedure is referred to as the {\em inverse problem}, i.e., given data that provides observations of the variable $I(t)$, namely $I_r(t)$, for some time points $t$ to identify the parameters vector $\th \in \Rb^{n+6}$ such that the mathematical model \eqref{mod-base} fits the data in the sense of the least-squares. More precisely, to solve the inverse problem of parameter estimation, we have to find the vector $\th \in \Rb^{n+6}$ that minimizes the sum of squares
\begin{equation}
\label{S}
S(\th)  = \frac 1{N^2} \SUM_{j=1}^M \big[ (I_r)_j - f(j,\th_2) I(j,\th_1) \big]^2\;.
\end{equation}
The objective function defined in \eqref{S} corresponds to the sum of squares of the absolute errors $\tfrac { (I_r)_j - f(j,\th_2) I(j,\th_1) }N$ relative to the size of the targeted population $N$, i.e., of the differences between the daily number of infected reported with symptoms, $(I_r)_j$, and the theoretical daily symptomatic infected $f(j,\th_2) I(j,\th_1)$, which is a fraction of the real symptomatic infected $I(j,\th_1)$, corresponding to the model solution evaluated at $(j,\th_1)$ at days $j=1,\cdots,M$, for a given value of the parameters vector $\th$, and where $M$ is the size of each dataset. The variable $I(t,\th_1)$ depends on $t$ and $\th_1=(\beta_1,\:\beta_2,\:\cdots,\: \beta_n,\:\gamma,\:\tau_1,\:\tau_2)\in \Rb^{n+3}$, while the fraction $f(t,\th_2)$ depends on $t$ and $\th_2=(a,\:k,\:I_m)\in \Rb^3$, with $\th=(\th_1,\:\th_2)\in \Rb^{n+6}$.

The minimum of the sum of squares $S(\th)$ is designated as $\hat \th$ for every dataset. The vector $\hat \th$ is called the {\em nonlinear least-squares estimator} denoted as {\em nonlinear LSE} hereafter. To minimize $S(\th)$, we applied the {\em Trust-Region Interior Reflective} (TIR) method implemented in Matlab\copyright\: as the subroutine {\em lsqnonlin}, specially adapted for solving nonlinear least-squares minimization problems.

Below, we describe the entire methodology extensively. In Subsection \ref{prac}, we provide details on the implementation of the lsqnonlin solver, specifically on the procedure to assess the quality of the optimal solution $\hat \th$, and on the stopping criterion associated with the algorithm, both concerning the convergence. Also, we computed the fit performance by some goodness-of-fit criteria and represented graphically the traditional sensitivity functions associated with the nonlinear LSE to assess parameter uncertainty, as explained in subsection \ref{fit-sen}.

\subsubsection{Parameter estimation}
\label{prac}

Convergence of TIR method, as described in \cite{Col-96}, is
theoretically achieved under general conditions. However, in practice,
the convergence depends strongly on the initial parameter estimations,
which has to be relatively close to the optimal solution. We were able to
efficiently minimize the objective function by a trial and error process,
by executing the codes hundreds of times, to guess the suitable initial
parameters vector that is different for every dataset
\cite{Cum-et-al-19,Cum-et-al-20}. It is worth noting that the initial
parameters that mostly influence the fitting results are
$(\beta_1,\:\beta_2,\:\cdots,\:\beta_n)$ that describe the mean rate of
COVID-19 transmission, according to the scenarios of expansion.
\textcolor{black}{As usual, there is a tradeoff among how well the model
outputs fit the empirical data, the number of parameters (increasing
$n$), and the risk of overfitting. 
Moreover, there is the human factor
involved in selecting the $n$ time intervals associated with each
parameter $\beta$, although changes in lockdown policies
determine reasonable interval boundaries. 
We will estimate parameters using
hand-picked intervals to address this issue, justifying our choices in
Section \ref{results}. In subsection \ref{ss:robust}, we will explore our
results' robustness using equal length intervals and varying values of
$n$.}

To evaluate the objective function $S(\th)$, we numerically solved the
model \eqref{mod-base} at days $j=1,\cdots, M$ for different parameters
vectors $\th$ depending on every dataset by applying a Rung-Kutta type
formulae \cite{L.F-2001}. We invoked the subroutine {\em dde23}
implemented in Matlab\copyright\:, designed for solving delay
differential equations (DDE) systems with constant time delays. In
addition, we had to reconstruct the function of history for the model
\eqref{mod-base} by interpolating the daily reported data of infected
with symptoms, recovered and dead considered for every dataset. The
interpolation used is a shape-preserving piecewise cubic as devised by
the {\em interp1} Matlab subroutine with the option {\em pchip}.

As for the options chosen for the lsqnonlin solver, we first choose the TIR method as the optimization algorithm, which is the default. Secondly, the maximum number of iterations was set at 1,000, while the maximum number of function evaluations was set at 20,000. Finally, the function tolerance was set at $1.0e-17$, whereas the norm of step tolerance, denoted by $tol$, was set variable between $1.0e-14$ and $1.0e-08$, depending on the targeted dataset. The norm of the step $\delta^{(p)}_{\th} >0$ measures the change between two successive iterates $\th^{(p)}$ and $\th^{(p+1)}$, which is defined as:
\begin{equation}
\label{step}
\delta^{(p)}_{\th} = \| \th^{(p+1)} - \th^{(p)} \|. 
\end{equation}
On the other hand,
\begin{equation}
\label{tol-fun}
\delta^{(p)}_S = \frac{ | S( \th^{(p)} ) - S( \th^{(p+1) } ) | } { 1+ | S( \th^{(p)} ) |  }\;,
\end{equation}
is the relative change in the sum of squares. The stopping criterion for our algorithm is defined as:
\begin{equation}
\label{stop}
\delta^{(p)}_S  < 1.0e-17 \quad \t{or} \quad \delta^{(p)}_{\th} < tol. 
\end{equation}
When the previous condition is met, the algorithm stops at iteration $p+1$ and returns an approximation of the nonlinear LSE ${\hat \theta}= \th^{(p+1)}$, as well as an approximation of the Jacobian matrix $\chi$ evaluated at $(j,\: \hat \theta)$ for every dataset; see Subsection \ref{fit-sen} for details.
To check the convergence of the algorithm, we also provided the first-order optimality measure, which measures how close the approximation is to the actual minimum of the sum of squares under the criterion of first-order partial derivatives. The first-order optimality measure is defined as the infinity norm of the gradient of the objective function evaluated at the nonlinear LSE $\hat \th$, i.e., the maximum absolute value of the partial derivatives of the objective function with respect to the $\th$ variables:
\begin{equation}
\label{opt-val}
\| \nabla S( \hat \theta) \|_{\infty} = \max_{\ell=1,\cdots,n+6} \left| \frac {\p S}{\p \th_{\ell}}(\hat \theta) \right| \;.
\end{equation}
Other metrics that quantify convergence are the norm of the step $\delta^{(p)}_{\th}$, the relative change in the sum of squares $\delta^{(p)}_S$ (see equations \eqref{step}-\eqref{tol-fun}), and the number of iterations $p+1$. Optionally, lsqnonlin displays the metrics of convergence at the end of its execution.

We computed the model parameters $\th=(\beta_1,\:\beta_2,\: \cdots, \: \beta_n, \: \gamma,\: \tau_1,\: \tau_2, \: a,\: k,\: I_m) \in \Rb^{n+6}$ within the following bounds: $1\leq \tau_1 \leq 14$, $7\leq \tau_2 \le 56$ days (the average for the time of incubation and removal, by recovery or death, is 5 and 14 days respectively; see \cite{China-WHO}), $0 \leq a \leq 1$, $\min \{ I_r(t) \} \leq I_m \leq \max\{ I_r(t)\} $ in the time interval considered, and all the rest of parameters are positive (i.e., between $0$ and $+\infty$).

\subsubsection{Goodness-of-fit criteria and parameter uncertainty}
\label{fit-sen}

We used statistical methods (from the context of nonlinear least-squares regression) to quantify the fit performance of the model to the data. More precisely, once estimating the nonlinear LSE $\hat \theta$, one may compute several goodness-of-fit criteria, which evaluate how well the model \eqref{mod-base} fits each dataset. We calculated $\hat \sigma^2$ and $\hat \sigma$, the unbiased variance of the residuals, and the root mean square error (RMSE) defined respectively in \cite{Banks, Greene} as
\begin{equation}
(\hat \sigma)^2 = \frac 1{M-n-6} \SUM_{j=1}^M \big[ (I_r)_j - f(j,\th_2) I(j,\th_1) \big]^2 = N^2\frac { S(\hat \theta) }{M-n-6}, \qquad \hat \sigma =N \sqrt{ \frac{ S( \hat \theta ) }{ M - n-6 } }\;. \label{est4.0}
\end{equation}
If the residuals $\tfrac{ \left[ (I_r)_j - f(j,\th_2) I(j,\th_1) \right]}N$ were normally distributed or if the datasets sizes $M$ were sufficiently large, then the estimated covariance matrix of the nonlinear LSE $\hat \theta$ would be expressed as
\begin{eqnarray}
Cov &=& \hat \sigma^2 \left[ \chi^t \chi \right]^{-1}, \quad \text{where}, \label{est2.1} \\[0.3cm]
\chi_{j,l} &=& \frac {\p (f  I)}{\p \th_l} (j,\:\hat \theta) \quad \t{for every } j=1,\cdots,M; \: \text{and every } l=1,\cdots,n+6\;. \label{est2.2} \\[0.3cm]
S_{\th_l}(t) &=& \chi_{t,l} \quad \t{for every } t;\: \t{and each } l=1,\cdots,n+6 \: \t{ fixed.} \label{sens}
\end{eqnarray}
The matrix $\chi$ is called the {\em traditional sensitivity function}, and it quantifies the variation of the state variable $f(t,\th_2) I(t,\th_1)$ with respect to changes in the parameters vector components $\th=(\th_1,\:\th_2)\in \Rb^{n+6}$. Sensitivity analysis can provide information about the relevance of data measurements to identify parameters; it then yields the basis for new tools to design inverse problem studies; see \cite{Banks-et-al08} for details. By using the covariance matrix given in \eqref{est2.1}, we can compute the standard error, $se$, and the normalized standard errors, $nse$, associated to the nonlinear LSE $\hat \theta$ from which we could quantify the accuracy of the parameter estimate. Both quantities are defined by
\begin{equation}
\label{est2}
se_l = \sqrt{ \left( Cov \right)_{l,l} }\;, \qquad nse_l = 100 \cdot \frac{ se_l }{ \hat \theta_l }\:, \quad \t{for every } l=1,\cdots,n+6.
\end{equation}
The quantities defined by \eqref{est2.1}-\eqref{est2} are asymptotic, i.e., they are valid only if each dataset size $M$ is sufficiently large because the nonlinear least-squares estimators are asymptotically normally distributed \cite{Banks, Greene}. We did not provide the $nse$ but plotted the sensitivities of parameters $\th_l$ with a $nse$ bigger than 100\%; see \eqref{sens}.

\section{Results and discussion}
\label{results}

In this section, we show and discuss the numerical results obtained according to the generalized SIR model with constant time delays \eqref{mod-base}, for each dataset corresponding to the selected regions of Chile for performing the calibration and prediction. We start by describing how we chose the different scenarios, followed by the fitting and prediction by using a figure that depicts the actual and calibrated infected, $I(t)$ and $f(t) I(t)$, respectively, computed according to our model. Next, we discuss parameter estimation's reliability by depicting the traditional sensitivity functions for those parameters having a $nse$ bigger than $100\%$ if it applies; see \eqref{sens}-\eqref{est2}. A table follows that shows quantitative measures to verify the convergence of the optimization algorithm. Concretely, the table shows the number of iterations made by the optimization solver to meet the stopping criterion, the sum of squares' optimal value, the RMSE, the first-order optimality measure, the norm of step, and the estimated parameters' values. Finally, we show numerical results to simulate an eventual change of scenario, which effect we measured by computing the relative errors for predicted data after the final time of calibration and discussing the number of infected' curve's long-time behavior.

We should warn that the actual values would depend on people behavior and of the measures taken by sanitary authorities of the state. Therefore, the results of the article should consider as a general trend on curves' behavior at simulation' conditions.

\subsection{Results for the Antofagasta Region}

Next, we show and discuss fitting and forecasting results for the Antofagasta Region, whose capital is Antofagasta, located 1093 km. to the North of Santiago of Chile straight line. The region has $N=691,854$ inhabitants, and it is representative of the country by its important ore activity. We fitted data starting from day 36 up to day 201, corresponding to April 7 and September 19, respectively (dataset size $M=165$). We chose April 7 as the initial time because the number of active accumulated cases was 69.

The region had four of nine communes under confinement (Antofagasta, Mejillones, Calama, and Tocopilla) in the following periods. Antofagasta and Mejillones from May 5 to May 29, and from June 23 to September 28; Calama from June 9 to September 21, and Tocopilla from June 23 to August 9. On average and taking into account the population density, we defined four scenarios for Antofagasta region as follows: $[t_0,t_1[=[36,64[$ for the early expansion free of quarantine (between April 7 and May 5); $[t_1,t_2[=[64,84[$ for the phase of fast propagation under the first quarantine (between May 5 and May 29); $[t_2,t_3[=[84,113[$ for a phase of quarantine' relaxation with relatively fast propagation (between May 29 and June 23); and $[t_3,t_4[=[113,201[$ for a phase of slowdown expansion under the second quarantine (between June 23 and September 19).

\begin{figure}[H]
\centering
\includegraphics[scale=0.75]{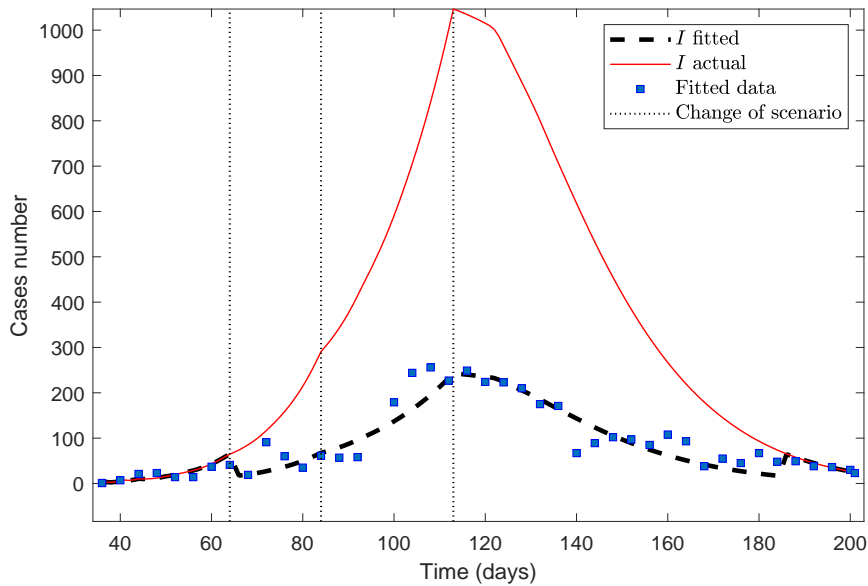}
\caption{Calibration for the Antofagasta Region dataset from April 7 up to September 19}
\label{fig1A}
\end{figure}

Figure \ref{fig1A} depicts the model fitted to the Antofagasta Region dataset, plotting them each four-time points.
Figure \ref{fig1A} shows that the calibrated and actual curves of infected are very different from each other, except in scenario 1 and near the end of scenario 4 of the pandemic. We expected this result since a few infected people would be almost all detected, so the calibrated and actual curves would coincide. On the other hand, we observe that the calibrated curve of infected fits quite well the data (infected reported $I_r(t)$). Indeed, since the dataset is large enough and encompasses the four stages of the pandemic, we estimated the parameters reliably. Only parameter $\hat k$, corresponding to the decay rate of fraction $f(t)$ (see \eqref{exp-fr}), has a $nse$ bigger than 100\%. Its sensitivity function varies in the interval [-1.2e-05, 7.8e-06] for all $t$ within the time interval of calibration; therefore, its estimate is unreliable. However, all the rest of the parameters have a $nse$ less than 100\%, so fitting is relatively reliable. Indeed, from Table \ref{tab1A}, one can observe that the overall fitting is suitable since the error is quite low (sum of squares and RMSE).
\begin{table}[H]
\centering
\begin{tabular}{|l|l|l|}
\hline
{\bf Quantity} & {\bf Symbol} & {\bf Value} \\ \hline
Iterations & $p+1$ & 37 \\ \hline
Sum of squares & $S(\hat \th)$ & $3.1149e-07$ \\ \hline
RMSE & $\hat \sigma$ & $31.0148$ \\ \hline
First-order optimality & $\| \nabla S(\hat \th) \|_{\infty}$ & $2.71e-06$ \\ \hline     
Norm of step & $\delta^{(p)}_{\th}$ & $6.6371e-08 $ \\ \hline
Mean transmission rate & $(\hat \beta_1,\: \hat \beta_2,\: \hat \beta_3,\: \hat \beta_4)$  & $(3.9591e-01,\:   3.0843e-01,\:   2.3631e-01,\:   1.6749e-01)$ \\ \hline
Mean removal rate & $\hat \gamma$ & $1.8455e-01$ \\ \hline
Mean time delays & $(\hat \tau_1, \: \hat \tau_2)$ & $(8.4047,\:   10.0222)$\\ \hline
Parameters of fraction $f(t)$ & $(\hat a,\: \hat k,\: \hat I_m)$ & $(23.20\%,\: 1.0243,\: 68.8445)$\\ \hline
\end{tabular}
\caption{\label{tab1A}Fit performance for the Antofagasta Region dataset from April 7 up to September 19}
\end{table}
The trend for the mean transmission rates is realistic. Indeed, $\hat \beta_1$ is the largest for scenario 1 (free of quarantine), followed by $\hat \beta_2$ for scenario 2 (with quarantine), followed by $\hat \beta_3$ for scenario 3 (free of quarantine), and finally $\hat\beta_4$ is the least of the four for scenario 4 (with quarantine). The same trend is observed for the fittings of the other datasets.

As explained in subsection \ref{sum}, we will show different predictions assuming that $\beta (t)$ may vary after September 19. Concretely, we run simulations by considering a mean transmission rate $\beta(t)$ defined by \eqref{beta-sim} taking $n=4$. The predicted data (not fitted) encompasses September 20 to October 05, $t_5=217$ that corresponds to this last date. Table \ref{tab2A} shows the range (minimum and maximum) and the average of the relative errors of the model' forecasting for different values of $\beta_5$, and where $t_4=210$ to try of capturing that none commune of the Antofagasta Region remained confined since September 28 (day 210).
\begin{table}[H]
\centering
\begin{tabular}{|l|l|l|l|}
\hline
{\bf Value}                               & {\bf Minimum} & {\bf Maximum} & {\bf Mean} \\ \hline
$\beta_5=2 \hat \beta_4$       &  $1.2416e-01$ & 1.3163 &  $4.3568e-01$ \\ \hline
$\beta_5=1.5\hat \beta_4$      & $1.4015e-01$ & 1.3163 &  $4.7957e-01$ \\ \hline
$\beta_5=3 \hat \beta_4$        & $1.7351e-02$ & 1.3163 &  $4.9284e-01$ \\ \hline
$\beta_5=1.25 \hat \beta_4$   & $1.3951e-01$ & 1.3163 &  $5.2653e-01$ \\ \hline
$\beta_5=1.125 \hat \beta_4$ & $1.2278e-01$ & 1.3163 &  $5.4590e-01$ \\ \hline
$\beta_5=\hat \beta_4$           & $1.4071e-01$ & 1.3163 &  $5.7411e-01$ \\ \hline
$\beta_5=0.75\hat \beta_4$    & $1.4134e-01$ & 1.3163 &  $6.2148e-01$ \\ \hline
$\beta_5=0.5  \hat \beta_4$    & $1.4198e-01$ & 1.3163 &  $6.6890e-01$ \\ \hline
$\beta_5=0.4  \hat \beta_4$    & $1.4227e-01$ & 1.3163 &  $6.8756e-01$ \\ \hline
$\beta_5=0.3  \hat \beta_4$    & $1.5134e-01$ & 1.3163 &  $7.0866e-01$ \\ \hline
$\beta_5=0.25  \hat \beta_4$  & $1.5275e-01$ & 1.3163 &   $7.1859e-01$ \\ \hline
$\beta_5=0.125 \hat \beta_4$ & $1.4228e-01$ & 1.3827 &   $7.3947e-01$ \\ \hline
\end{tabular}
\caption{\label{tab2A}Predictions of the infected curve under different scenarios for the Antofagasta Region}
\end{table}
From Table \ref{tab2A}, we observe that $\beta_5=2 \hat \beta_4$ yields the least relative error mean, which implies that after the quarantine finished on September 28, the mean rate of transmission doubled, producing a negative change of scenario. Our model forecasts that, if these conditions kept, a COVID-19 second-outbreak would occur in the Antofagasta Region. The peak of this new outbreak would occur around January 24, 2021, with an estimate of 102,154 real infected and 23,703 daily reported infected.

\subsection{Results for the Metropolitan Region}
\label{ss:results_MR}

Next, we show and discuss the results of fitting and forecasting for the
Metropolitan Region (RM), whose capital is Santiago of Chile, also the
capital of Chile. We fitted data starting from day 15 until day 203
corresponding to March 17 and September 21, respectively (dataset size
$M$=189). We chose March 17 as the starting day because the number of newly
infected with symptoms was 29, with 152 active accumulated cases.

The region has $N=8,125,072$ inhabitants (around 41.76\% of the total
inhabitants of Chile) and 52 communes. From these, 47 communes have been
under selective quarantine at different periods (few of which are yet),
the most of which on average between May 7 and June 30. Therefore, we
defined the three scenarios for MR as follows: $[t_0,t_1[=[15,66[$ for
the first stage that encompasses the period from March 17 to May 7 (early
expansion free of quarantine on average); $[t_1,t_2[=[66,120[$ for the
second phase that ranges from May 7 to June 30 (fast propagation under
quarantine); and $[t_2,t_3[=[120,203[$ for the third step that goes from
June 30 up to September 21 (slowdown expansion with relaxation of
quarantine on average).

\begin{figure}[H]
\centering
\includegraphics[scale=0.75]{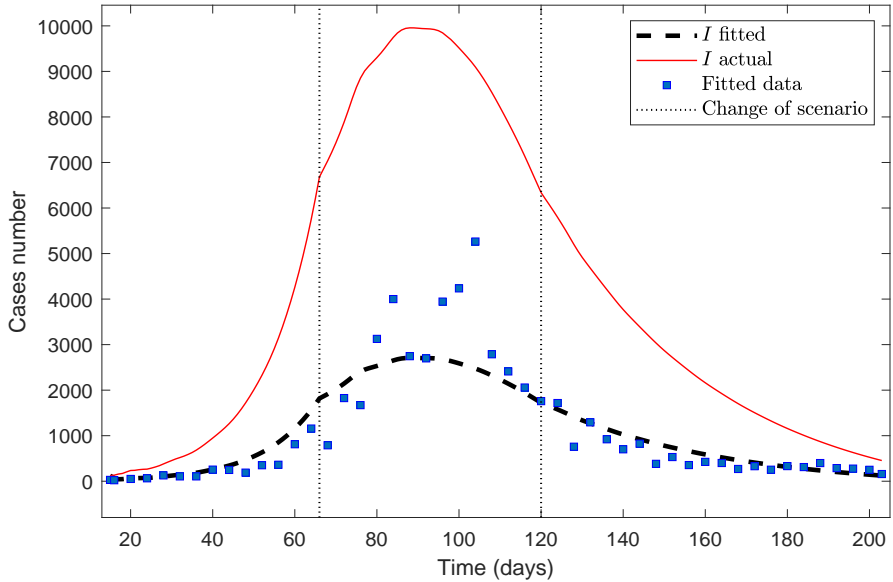}
\caption{Calibration for MR dataset from March 17 up to September 21}
\label{fig1}
\end{figure}

Figure \ref{fig1} depicts the model fitted to the MR dataset, plotting
them each four-time points.
Figure \ref{fig1} shows that curves of calibrated and actual infected are
very different from each other. On the other hand, we observe a good
calibration of the model to the data since the computed curve of
infected, $f(t)I(t)$, fits quite well the daily reported infected,
$I_r(t)$. Indeed, since the dataset is large enough and encompasses the
three stages of the pandemic, we estimated the parameters reliably, which
is reinforced by the fact that the maximum $nse$ is $4.4e-04$\%, meaning
very accurate calibration results. To further appreciate the suitable
fitting of the model, Table \ref{tab1} presents its performance.
\begin{table}[H]
\centering
\begin{tabular}{|l|l|l|}
\hline
{\bf Quantity} & {\bf Symbol} & {\bf Value} \\ \hline
Iterations & $p+1$ & 32 \\ \hline
Sum of squares & $S(\hat \th)$ & $9.3320e-07$ \\ \hline
RMSE & $\hat \sigma$ & $589.9672$ \\ \hline
First-order optimality & $\| \nabla S(\hat \th)\|_{\infty}$ & $2.98e-05$ \\ \hline
Norm of step & $\delta^{(p)}_{\th}$ & $6.1739e-08$ \\ \hline
Mean transmission rate & $(\hat \beta_1,\: \hat \beta_2,\: \hat \beta_3)$  & $(2.7229e-01,\:   1.7869e-01,\:   1.8942e-01)$ \\ \hline
Mean removal rate & $\hat \gamma$ & $1.8681e-01$ \\ \hline
Mean time delays & $(\hat \tau_1, \: \hat \tau_2)$ & $(7.7464,\:   11.9062)$\\ \hline
Parameters of fraction $f(t)$ & $(\hat a,\: \hat k,\: \hat I_m)$ & $(27.21\%,\:  4.5644e-01,\:   131.1859)$\\ \hline
\end{tabular}
\caption{\label{tab1}Fit performance for the MR dataset from March 17 up to September 21}
\end{table}
As explained in subsection \ref{sum}, we will show different predictions assuming that $\beta (t)$ may vary after September 21. Concretely, we run simulations by considering a mean rate of transmission $\beta(t)$ defined by \eqref{beta-sim} for $n=3$. The predicted data (not fitted) encompasses September 22 to 28, $t_4=210$ that corresponds to this last date. Table \ref{tab2} shows the range (minimum and maximum) and the average of the relative errors of the model' forecasting for different values of $\beta_4$ and
for $t_3=182$ to try of capturing that almost none commune of the MR remained confined since August 31 (day 182).
\begin{table}[H]
\centering
\begin{tabular}{|l|l|l|l|}
\hline
{\bf Value} & {\bf Minimum} & {\bf Maximum} & {\bf Mean} \\ \hline
$\beta_4=1.125 \hat \beta_3$ & $1.1883e-01$ & $9.2232e-01$ & $4.6393e-01$ \\ \hline
$\beta_4=\hat \beta_3$         &  $3.7086e-01$ & $7.3887e-01$ & $6.1924e-01$ \\ \hline
$\beta_4=1.25\beta_3$         & $1.2109$ &   $2.4539$ &  $1.8206$ \\ \hline
$\beta_4=0.125 \hat \beta_3$ & $6.6810e-01$ & $9.5068$ & $4.2873$ \\ \hline
$\beta_4=0.75\beta_3$   &  $3.9652$ &  $5.6334$ & $4.7641$ \\ \hline
$\beta_4=1.5\beta_3$    &   $4.0958$ & $6.7886$ & $5.3415$ \\ \hline
$\beta_4=0.25  \hat \beta_3$ & $3.1006$ & $1.0145e+01$ & $5.6799$ \\ \hline
$\beta_4=0.4  \hat \beta_3$  &  $4.8842$ & $9.9985$ & $6.5442$ \\ \hline
$\beta_4=0.5  \hat \beta_3$  &  $5.0977$ & $9.3263$ & $6.6037$ \\ \hline
$\beta_4=2 \hat \beta_3$    &  $1.2883e+01$ & $2.2141e+01$ & $1.6301e+01$ \\ \hline
$\beta_4=3 \hat \beta_3$    &  $4.3485e+01$ & $8.4643e+01$ & $5.7122e+01$ \\ \hline
\end{tabular}
\caption{\label{tab2}Predictions of the infected curve under different scenarios for the MR}
\end{table}
From Table \ref{tab2}, we observe that $\beta_4=1.125\hat
\beta_3$  yields the least relative error (minimum and mean), which
reinforces the assumption that the quarantine is valid for the MR between
May 7 and June 30, on average, and that there was a little change of
scenario from August 31 until September 28. Our model forecasts that, if
these conditions kept, a COVID-19 second-outbreak would occur in the MR.
The peak of this new outbreak would occur around February 27, 2021, with
an estimate of 154,698 real infected and 42,086 daily reported infected.

\subsection{Results for the Ñuble Region}

Next, we show and discuss fitting and forecasting results for the Ñuble
Region, located 403.8 km. to the south of Santiago of Chile, and whose
capital city is Chillán (population $N=511,551$). On September 21,
Chillán was the fourth city with more active infected per 100,000
inhabitants (407 cases). The three scenarios considered here are more
easily identifiable than for the MR and the Antofagasta Region since the
quarantines in the Ñuble Region were imposed mostly in Chillán at a
single date within the time interval of calibration.

We split the results into two parts; first, we did the fitting for the
data from March 21 to April 27 and next from March 21 to September 02.
Therefore, the scenarios are the following: $[t_0,t_1[=[19,28[$ for the
first stage that encompasses the period from March 21 up to March 30
(early expansion); $[t_1,t_2[=[28,52[$ for the second phase that ranges
from March 30 up to April 23 (fast propagation under quarantine); and
$[t_2,t_3[$ where $t_2=52$ and $t_3=57$ or $t_3=184$ for the third step
that goes from April 23 to April 27 and to September 2 (slowdown
expansion, free of quarantine), corresponding to the first and second
result, respectively.

\subsubsection{Results for fitting until April 27}

To compare with the only quantitative results we know of for the fitting of data from the Ñuble Region (see the first report by Cabrera-Vives et al. \cite{Cabrera}), we did the fitting starting from day 19 (March 21) until day 57 (April 27) (dataset size $M=39$), when the accumulated infected at the Ñuble Region were 58. It is worth noting that, because of the small number of data points, the third scenario does not represent a spreading's slowdown. We performed this fitting only to compare our results with that of Cabrera-Vives and collaborators' first report.

\begin{figure}[H]
\centering
\includegraphics[scale=0.7]{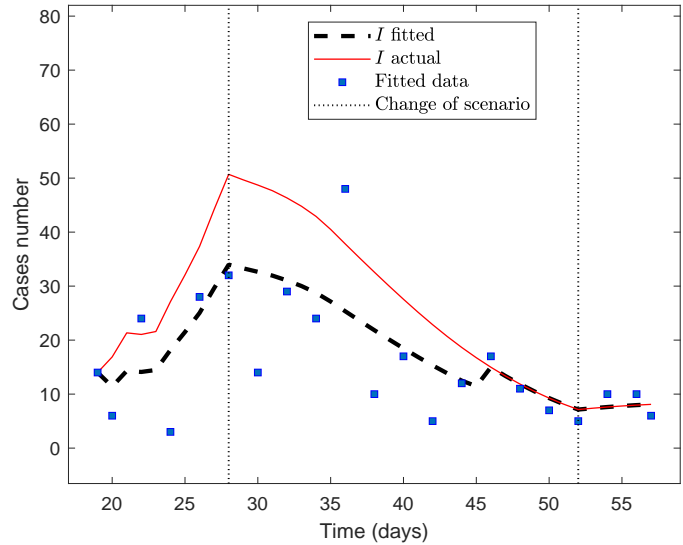}
\caption{Calibration for the Ñuble Region dataset from March 21 up to April 27}
\label{fig3}
\end{figure}

\begin{figure}[H]
\centering
\includegraphics[scale=0.6]{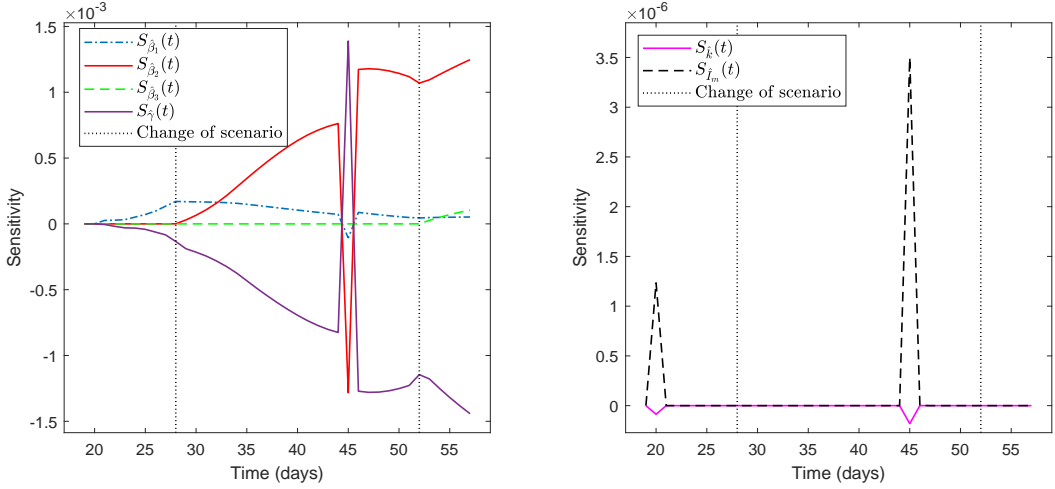}
\caption{Sensitivity functions for parameters estimated with a nse bigger than 100\%}
\label{fig4}
\end{figure}

Figure \ref{fig3} depicts the model fitted to the dataset for the Ñuble
Region, plotting the data each two-time points.
From Figure \ref{fig3}, we observe that the calibrated curve of infected does not fit suitably the data, despite that the optimization solver converged (the last change in the relative sum of squares is less than the value of the function tolerance). Indeed, the calibrated curve does not represent the trend of not fitted data from April 28 to May 02 (not reported here). Following the findings in \cite{Banks-et-al08}, the explanation is that the data used to make the fitting is close to the zero equilibrium (the early stage of the pandemic), suggesting limited ability to determine some parameters reliably. This is related to the traditional sensitivity functions, depicted in Figure \ref{fig4}, for the parameters estimated with a $nse$ bigger than 100\%; see \eqref{sens}-\eqref{est2}.
From Figure \ref{fig4}(right), we observe that parameters $\hat k$ and
$\hat I_m$ (see equation \eqref{exp-fr}) have the smallest sensitivities
(very near to zero for every $t$), which implies that their estimations
are not reliable. By contrast, Figure \ref{fig4}(left) shows that
parameter $\hat \gamma$ has the largest sensitivity, which means that its
estimate is relatively better than the other parameters having a $nse$
bigger than 100\%. In the case of parameters $(\hat \beta_1,\: \hat \beta
_2,\: \hat \beta_3)$, $S_{\hat \beta_1}(t)$ and $S_{\hat \beta_3}(t)$ are
small in the regions of scenarios 1 and 3, where $\hat \beta_1$ and $\hat
\beta_3$ represent the mean rates of transmission, respectively; by
contrast, $S_{\hat \beta_2}(t)$ is large enough in the region of scenario
2, where $\hat \beta_2$ describes the mean rate of transmission.
Consequently, the estimation of $\hat \beta_2$ is relatively better than
those of $\hat \beta_1$ and $\hat \beta_3$. We will see that fitting the
data in the three phases of the pandemic provides better results. 

Table \ref{tab3} shows the quantitative results of the fitting.
\begin{table}[H]
\centering
\begin{tabular}{|l|l|l|}
\hline
{\bf Quantity} & {\bf Symbol} & {\bf Value} \\ \hline
Iterations & $p+1$ & 41 \\ \hline
Sum of squares & $S(\hat \th)$ & $ 2.0949e-08$ \\\hline
RMSE & $\hat \sigma$ & $13.5179$ \\ \hline
First-order optimality & $\|\nabla S(\hat \th)\|_{\infty}$ & $ 2.19e-08 $ \\ \hline
Norm of step & $\delta^{(p)}_{\th}$ & $1.0606e-10$ \\ \hline
Mean transmission rate & $(\hat \beta_1,\:\hat \beta_2,\:\hat \beta_3)$  & $(3.7849e-01,\:   2.2784e-02,\:   1.1015e-01  )$ \\ \hline
Mean removal rate & $\hat \gamma$ & $7.5470e-02$ \\ \hline
Mean time delays & $(\hat \tau_1, \: \hat \tau_2)$ & $(5.4927,\:  7.1330 )$\\ \hline  
Parameters of fraction $f(t)$ & $(\hat a,\: \hat k,\: \hat I_m)$ & $(67.02\%,\:   7.7127,\:  16.3431)$\\ \hline
\end{tabular}
\caption{\label{tab3}Fit performance for the Ñuble Region dataset from March 21 up to April 27}
\end{table}
The results are comparable with that from the first report by
Cabrera-Vives et al. \cite{Cabrera} (a similar RMSE), although we did the
fitting by using the real date at which the authorities imposed the
quarantine. Our model also forecasts that, if the same conditions of the
third scenario kept, the outbreak's peak would occur around February 20,
2021, that qualitatively coincides with the prediction obtained by
Cabrera-Vives et al. \cite{Cabrera}; see the bottom of Figure 1 from the
first report. However, according to the previous discussion, the fitting
obtained from data near an equilibrium point produces wrong prediction
results \cite{Banks-et-al08, Banks-et-al07}.

\subsubsection{Results until September 2}

\begin{figure}[H]
\centering
\includegraphics[scale=0.75]{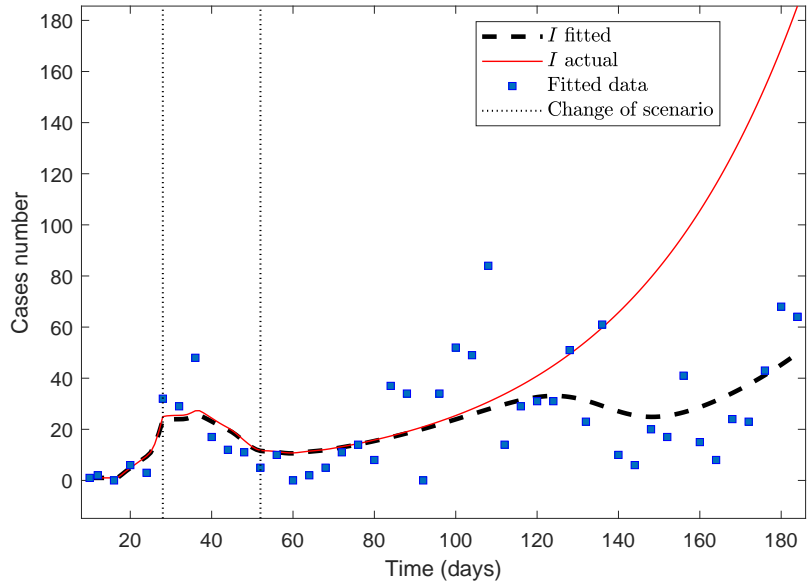}
\caption{Calibration for the Ñuble Region dataset from March 21 up to September 2}
\label{fig6}
\end{figure}

\begin{figure}[H]
\centering
\includegraphics[scale=0.7]{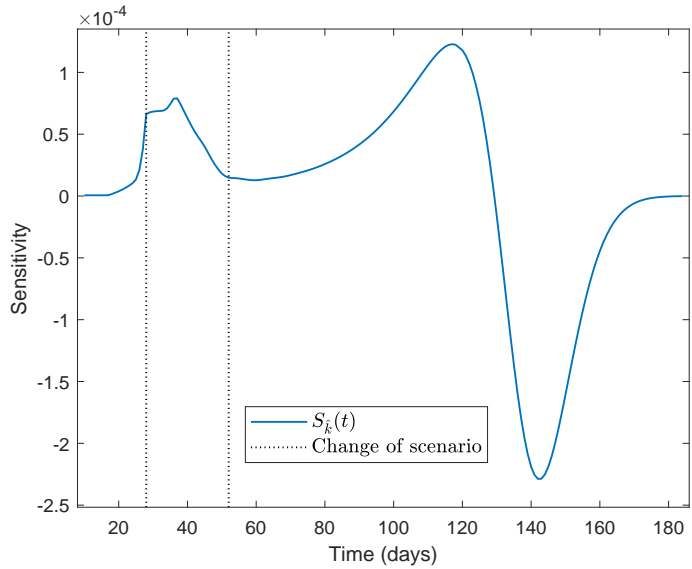}
\caption{Sensitivity function for parameter $\hat k$}
\label{fig7}
\end{figure}

Now, we show the results of fitting and predictions for the data from
Ñuble Region, widening the time window from March 12 to September 02
(dataset size $ M = 175 $), considering the same constraints as before
(quarantine from March 30 to April 23), but now $ t_3 = 184 $ instead of
57 (day 184 corresponds to September 2). Figure \ref{fig6} depicts the
model fitted to the dataset for the Ñuble Region, plotting the data at
each four-time points.
From Figure \ref{fig6}, we observe that the calibrated curve of infected
fits quite well the data. In this case, since the dataset is large enough
and encompasses data from the three stages of the pandemic, this time, we
estimated the parameters reliably, except maybe for $\hat k$ that is the
only with a $nse$ bigger than 100\%. Figure \ref{fig7} depicts the
sensitivity function of $\hat k$.
From Figure \ref{fig7}, we observe that $S_{\hat k}(t)$ is relatively
large in the region corresponding to scenario 3, and therefore estimate of $\hat k$ is relatively reliable. The result is better than the last
fitting (data until April 27); see Figure \ref{fig4} and the discussion
below it. 

Table \ref{tab4} shows the corresponding quantitative results of the fitting.
\begin{table}[H]
\centering
\begin{tabular}{|l|l|l|}
\hline
{\bf Quantity} & {\bf Symbol} & {\bf Value} \\ \hline
Iterations & $p+1$ & 68 \\ \hline
Sum of squares & $S(\hat \th)$ & $ 1.3061e-07 $ \\ \hline
RMSE & $\hat \sigma$ & $14.4360$ \\ \hline
First-order optimality & $\| \nabla S(\hat \th) \|_{\infty}$ & $ 3.65e-07 $ \\ \hline
Norm of step & $\delta^{(p)}_{\th}$ & $1.1492e-08$ \\ \hline
Mean transmission rate & $(\hat \beta_1,\:\hat \beta_2,\:\hat \beta_3)$  & $(1.3248,\:   1.3487e-01,\:   1.9592e-01)$ \\ \hline
Mean removal rate & $\hat \gamma$ & $1.7578e-01$ \\ \hline
Mean time delays & $(\hat \tau_1, \:\hat  \tau_2)$ & $(7.4305,\:  9.4810)$\\ \hline  
Parameters of fraction $f(t)$ & $(\hat a,\:\hat k,\:\hat I_m)$ & $(26.78\%,\: 9.5539e-02,\:   50.9513)$\\ \hline
\end{tabular}
\caption{\label{tab4}Fit performance for the Ñuble Region dataset from 12 March to 02 September}
\end{table}
Consistently with the results obtained fitting the dataset until April
27, our model forecasts that, if the same conditions of the third
scenario kept, the outbreak's peak would occur around March 02, 2021.
However, we would expect a change of scenario since September 2 because
authorities re-imposed confinement in Chillán due to the high increase in
daily new cases.

As explained in subsection \ref{sum}, we will show different predictions assuming that $\beta (t)$ may vary after September 2 due to the confinement (helpful or counterproductive measure depending on people behavior). Concretely, we run simulations by defining $\beta(t)$ as in \eqref{beta-sim} (with $n=3$) to predict the data (not
fitted) between September 3 and September 22 (data number $=20$), and
putting $t_3=184$ and $t_4=204$ that correspond to September 2 and 22, respectively. Table
\ref{tab5} shows the range (minimum and maximum) and the average of the
relative errors of the model' forecasting for different values of
$\beta_4$.
\begin{table}[H]
\centering
\begin{tabular}{|l|l|l|l|}
\hline
{\bf Value} & {\bf Minimum} & {\bf Maximum} & {\bf Mean} \\ \hline
$\beta_4=0.75 \hat \beta_3$   &  $1.8775e-02$  & $1.0677$ &  $3.0481e-01$ \\ \hline
$\beta_4=0.5   \hat \beta_3$   &  $2.6869e-02$ &  $2.9773$ &  $7.2569e-01$ \\ \hline
$\beta_4= \hat \beta_3$          &  $5.2774e-02$  & $2.8704$  & $7.9289e-01$ \\ \hline
$\beta_4=1.125 \hat \beta_3$ &  $5.7721e-03$  & $3.8428$   & $1.1393$ \\ \hline
$\beta_4=0.4  \hat \beta_3$    &  $6.0453e-02$   &  $4.8728$ & $1.2234$ \\ \hline
$\beta_4=1.25 \hat \beta_3$   &  $2.1683e-02$ &  $5.0685$  & $1.5241$ \\ \hline
$\beta_4=0.25  \hat \beta_3$  &  $2.6635e-02$  & $7.1286$  & $2.0765$ \\ \hline
$\beta_4=1.5 \hat \beta_3$     &  $1.0001e-01$  & $8.2770$  & $2.3404$ \\ \hline
$\beta_4=0.125 \hat \beta_3$ &  $1.0034e-02$  & $8.4642$  & $2.7584$ \\ \hline
$\beta_4=3 \hat \beta_3$       &  $2.2141e-01$  & $3.9640e+01$  & $9.1426$ \\ \hline
\end{tabular}
\caption{\label{tab5}Predictions of the infected curve under different scenarios for the Ñuble Region}
\end{table}
From Table \ref{tab5}, we observe that $\beta_4=0.75 \hat \beta_3$ yields
the least relative error maximum and mean. The interpretation is that
quarantine effectively reduced the transmission rate, and therefore there
was a positive change of scenario since September 2. Our model forecasts
that, if these conditions kept, a COVID-19 second-outbreak would not
occur in the Ñuble Region.

\begin{figure}[H]
\centering
\includegraphics[scale=0.7]{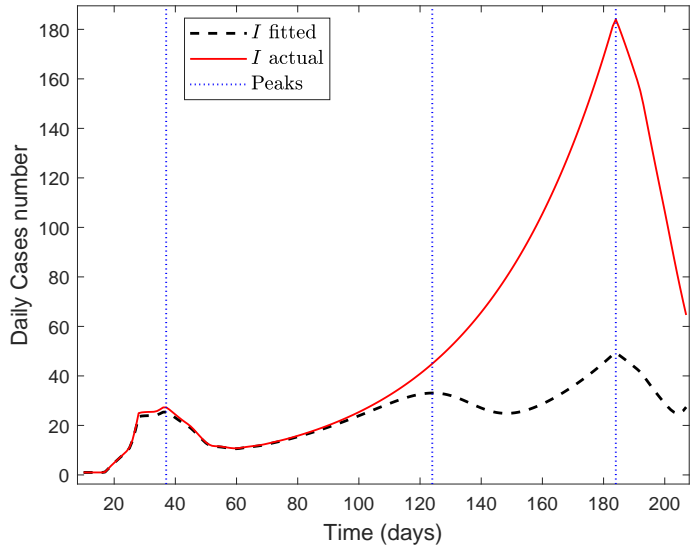}
\caption{Forecasted curve of infected for Ñuble Region from March 21 to September 25}
\label{fig8}
\end{figure}

Finally, Figure \ref{fig8} depicts the calibrated and real curves of infected for the Ñuble Region until September 25, by taking $\beta_4=0.75 \hat \beta_3$ as before. The third peak corresponds to September 2, the final calibration date.
The result of Figure \ref{fig8} coincides with a report by the Chilean Society of Intensive Medicine, published on September 25, which pointed out that Chillán underwent the third peak of the COVID-19 outbreak by this date; see \cite{rep}.

\subsection{On the robustness of the optimization algorithm and the estimation of the infection fatality rate}
\label{ss:robust}

To assess the robustness of the $\beta$ parameter estimations we repeated
the fitting processed we described in Subsection \ref{ss:results_MR}
after modifying some assumtions. Instead of assuming that $\beta(t)$ is
piecewise constant and defining the three scenarios based on actual
policy changes, here we considered $\beta(t)$ to be piecewise linear with
$n-1$ equal length time intervals. More formally we replaced equation
\ref{beta} with
\begin{equation}
\beta(t) = \sum_{i=1}^n \beta_i \textbf{T}_i(t)
\end{equation}
where $\textbf{T}_i$ is the tent function with value of one at $t_i$ and
tapers off to zero at $t_{i-1}$ and $t_{i+1}$. Thus, we determine $n$
values for the $\hat{\beta}_i$'s. We tried 5 values for $n$ and we
summarize the results in Table \ref{t:robust}.
\begin{table}[H]
\centering
\begin{tabular}{|c|c|c|c|c|c|c|c|c|}
     \hline
$n$ & $\hat{\beta}_1$ & $\hat{\beta}_2$ & $\hat{\beta}_3$ & $\hat{\beta}_4$ & $\hat{\beta}_5$ & $\hat{\beta}_6$ & $\hat{\beta}_7$  & $\hat{\beta}_8$\\
     \hline
8 & $7.5e-01$ & $8.2e-02$ & $2.1e-01$ & $1.4e-01$ & $1.8e-01$ & $2.1e-01$ & $1.2e-01$ & $4.2e-01$\\ \hline
7 & $1.3e-01$ & $1.6e-01$ & $1.7e-01$ & $1.6e-01$ & $9.2e-02$ & $5.8e-02$ & $2.4e-01$\\  \cline{1-8}
6 & $1.7e-01$ & $3.3e-01$ & $1.3e-01$ & $1.7e-01$ & $2.2e-01$ & $8.7e-02$ \\
     \cline{1-7}
5 & $5.3e-01$ & $1.9e-01$ & $1.4e-01$ & $1.9e-01$ & $9.0e-02$\\  \cline{1-6}
4 & $1.7e-01$ & $1.5e-01$ & $1.3e-01$ & $3.9e-01$\\ \cline{1-5}
\end{tabular}
\caption{\label{t:robust} Estimators for $\beta(t)$ using $n-1$ segments
piecewise linear functions. }
\end{table}
The results in Table \ref{t:robust} are, up to a point, compatible with
the values shown in Table \ref{tab1}. For example, we can see that for
values of $i\approx n/2$ the $\hat{\beta}_i$'s are in the [0.13, 0.19]
range consistently for $\hat \beta_2$ from Table \ref{tab1}. Moreover,
for $n=4$ and $n=5$ the same trend follows as in Table \ref{tab1}, i.e.,
the values for $\hat{\beta}_i$ are relatively large for both small and
large $i$'s (first and last scenario). We already stated in Section
\ref{results} that this was a consequence of the different scenarios
considered for each dataset. The same trend is not seen if $n=7$. This is
natural, as having many time  intervals, and therefore many
$\hat{\beta}_i$'s parameters to adjust will tend to cause overfitting.
Indeed, we observe across rows of Table \ref{t:robust} a relatively large
variability (range [0.13, 0.75]) among the $\hat{\beta}_1$'s. The same
can be said about the values for $\hat{\beta}_n$ (range $[0.09, 0.42]$).
One consequence of  this observation is that predictions are likely to be
very inaccurate if we simply assume that $\hat{\beta}_n$ will hold during
an extrapolation beyond the last data point after trying the fitting with
a single $n$. Nonetheless, we can use the ranges the $\hat{\beta}_n$'s
lie on for a scenario analysis as we did in Section \ref{results}
(tables \ref{tab2A}, \ref{tab2} and \ref{tab5}). Another consequence is
the difficulty to estimate precisely the  effectiveness of the quarantine
procedures. This follows from the variability within each row of Table
\ref{t:robust}. In principle, quarantines should lower the values for
$\beta$. However, because of the inherent inaccuracy of the estimation
during the initial disease propagation, and the enforcement of
quarantines as soon as the cases multiply (and thus the estimation
becomes reliable), computing the temporal changes in $\beta$ accurately
is more than challenging. Yet, for small values of $n$ (say 4 or 5), the
qualitative behavior of the $\beta_i$'s we expect is observed.

Furthermore, it is of practical interest to estimate the lethality of a
disease. The {\em infection fatality rate} (IFR), the probability of
dying for an infected person, is computed as the ratio of dead subjects
to the number of infected people for the targeted population \cite{Ioan}.
However, according to our model, there is a time delay between that an
infected person recovers or dies ($\tau_2$), and therefore to compute the
IFR at a given day $t$, we divided $D(t)$ (the number of death at $t$)
into the number of infected at time $t-\tau_2$. To exemplify the model's
usefulness, we will show IFR estimations computed for each dataset by
combining empirical data and model outputs. To do that, we will use
$\tau_2$ from Tables \ref{tab1A}, \ref{tab1} and \ref{tab4}, $D(t)$ the
confirmed number of deaths for each region reported in \cite{data-reg},
and the actual number of infected $I(t)$ computed from the outputs of
model \ref{mod-base}. As usual, we divided the process into the
scenarios, as described in Section \ref{results}, plus the pandemic's
entire period used in the calibration. Table \ref{ifr} shows the IFR
medians in percentage.
\begin{table}[H]
\centering
\begin{tabular}{|l|l|l|l|l|l|}
\hline
{\bf Region} & {\bf Scenario 1} & {\bf  Scenario 2} & {\bf  Scenario 3} & {\bf Scenario 4} & {\bf Entire} \\ \hline
Antofagasta & 0 &  0  &  6.8758e-01 &  8.8174e-01 &  7.9727e-01 \\ \hline
Metropolitan & 4.2424e-01        & 5.0989e-01         &   1.2302   & Does not apply   & 9.6585e-01 \\ \hline
Ñuble &  0  &   0 &    0 & Does not apply &   0 \\ \hline
\end{tabular}
\caption{\label{ifr}IFR median estimation for each dataset}
\end{table}
We computed the median since it is more appropriate for skewed
distribution, such as the confirmed number of deaths, $D(t)$, which are
raw data. We did not plot the IFR since it presents many oscillations and
a few outlier values because of uncertainty in data and the numerical
error of calibration results. The values of Table \ref{ifr} are
consistent with the recent results in \cite{Ioan}, where the author
claims that $\text{IFR} \in [0\%, 1.54\%]$, which implies that the curve
of real infected provided by the model is accurate since it yields a
realistic value for the IFR's median in the case of MR. By contrast, for
the Ñuble and Antofagasta Regions, the confirmed death cases are not
large enough to yield an accurate IFR's estimation (for neither mean nor
median).

\section{Conclusions}

We described and successfully implemented a clear, efficient, and reproducible parameter estimation methodo\-logy to a generalized SIR model with constant time delays that can reproduce complex dynamics for COVID-19. We illustrated our methodology and modeling by carrying out parameter estimation for three datasets corresponding to Chile's three representative regions, although this can be applied to any country. From a methodological viewpoint, we assessed the reliability of estimated parameters and shown that when the data are located in the transition from the zero to the non-zero equilibria (encompassing all the stages of pandemic spread), the parameters are reliably estimated. Also, we verified our optimization methodology's robustness by considering an arbitrary number of $ n $ scenarios.

The numerical results allow us to forecast the general trend of the infected' curve, the calibrated and the real, and provide some predictions that allow us to prognosticate a possible COVID-19 second-outbreak. This prognosis is valid only if the conditions that allowed arrive at the estimated mean transmission rate are kept. In the absence of an effective vaccine or drug, these conditions are essentially translated into self-care.  Finally, we showed that our model is precise enough to reproduce the infection fatality rate fairly accurately, according to the reported dead cases.

There are two lines of research that we would like to tackle. First, we expect to apply our model and methodology to other regions/communes in Chile. Because of the heterogeneity among their socio-economic conditions, the pandemic has affected them differently. On the other hand, we intend to calibrate our model to the reported dead to estimate more accurately and quantitatively the infection fatality rate of COVID-19, including structures of age and the dynamics of interactions among individuals of diverse age groups.

\section*{Funding}
The Centre for Biotechnology and Bioengineering (CeBiB) supported this
work under PIA grant FB-01 from ANID. P.C.'s work was also supported by
DIUBB 193409 3/R regular research project from the University of Bío-Bío.

\section*{Declaration of competing interest}
The authors declare that they have no conflict of interest.

\bibliographystyle{elsarticle-num}
\bibliography{covid-19-ref}      

\begin{thebibliography}{10}
\expandafter\ifx\csname url\endcsname\relax
  \def\url#1{\texttt{#1}}\fi
\expandafter\ifx\csname urlprefix\endcsname\relax\def\urlprefix{URL }\fi
\expandafter\ifx\csname href\endcsname\relax
  \def\href#1#2{#2} \def\path#1{#1}\fi

\bibitem{CMM}
A.~Cancino, P.~Gajardo, R.~Lecaros, C.~Mu\~noz, H.~Ram\'irez, J.~Ortega,
  Covid-19 en chile, http://www.cmm.uchile.cl/?p=37663 (2020).

\bibitem{Cabrera}
G.~Cabrera-Vives, C.~Donoso-Oliva, M.~Mart\'inez, R.~Molina, A.~S\'anchez,
  Informe proyecciones covid-19 udec,
  https://github.com/guille-c/Covid-19/tree/master/Informes (2020).

\bibitem{candia}
M.~Candia~Reyes, G.~Vergara-Hermosilla,
  \href{https://hal.archives-ouvertes.fr/hal-02560526}{{Estimaci{\'o}n de casos
  no reportados de infectados de COVID-19 en Chile, el Maule y la Araucan{\'i}a
  durante marzo de 2020}}, working paper or preprint (May 2020).
\newline\urlprefix\url{https://hal.archives-ouvertes.fr/hal-02560526}

\bibitem{G-N}
C.~Guerrero-Nancuante, R.~{Manr{\'{i}}quez P}, {An epidemiological forecast of
  COVID-19 in Chile based on the generalized SEIR model and the concept of
  recovered}, Medwave 20~(04) (2020) e7898--e7898.
\newblock \href {https://doi.org/10.5867/medwave.2020.04.7898}
  {\path{doi:10.5867/medwave.2020.04.7898}}.

\bibitem{Peng}
L.~Peng, W.~Yang, D.~Zhang, C.~Zhuge, L.~Hong,
  \href{https://www.medrxiv.org/content/early/2020/02/18/2020.02.16.20023465}{Epidemic
  analysis of covid-19 in china by dynamical modeling}, medRxiv (2020).
\newblock \href
  {http://arxiv.org/abs/https://www.medrxiv.org/content/early/2020/02/18/2020.02.16.20023465.full.pdf}
  {\path{arXiv:https://www.medrxiv.org/content/early/2020/02/18/2020.02.16.20023465.full.pdf}},
  \href {https://doi.org/10.1101/2020.02.16.20023465}
  {\path{doi:10.1101/2020.02.16.20023465}}.
\newline\urlprefix\url{https://www.medrxiv.org/content/early/2020/02/18/2020.02.16.20023465}

\bibitem{Cum-et-al-19}
P.~Cumsille, M.~Godoy, Z.~P. Gerdtzen, C.~Conca,
  \href{https://doi.org/10.1371/journal.pone.0217332}{Parameter estimation and
  mathematical modeling for the quantitative description of therapy failure due
  to drug resistance in gastrointestinal stromal tumor metastasis to the
  liver}, PLOS ONE 14~(5) (2019) 1--27.
\newblock \href {https://doi.org/10.1371/journal.pone.0217332}
  {\path{doi:10.1371/journal.pone.0217332}}.
\newline\urlprefix\url{https://doi.org/10.1371/journal.pone.0217332}

\bibitem{Cum-et-al-20}
G.~Badillo, P.~Cumsille, L.~Segura-Ponce, G.~Pataro, G.~Ferrari,
  \href{https://doi.org/10.1080/17415977.2020.1717488}{An efficient
  optimization methodology of respiration rate parameters coupled with
  transport properties in mass balances to describe modified atmosphere
  packaging systems}, Inverse Problems in Science and Engineering 28~(10)
  (2020) 1361--1383.
\newblock \href
  {http://arxiv.org/abs/https://doi.org/10.1080/17415977.2020.1717488}
  {\path{arXiv:https://doi.org/10.1080/17415977.2020.1717488}}, \href
  {https://doi.org/10.1080/17415977.2020.1717488}
  {\path{doi:10.1080/17415977.2020.1717488}}.
\newline\urlprefix\url{https://doi.org/10.1080/17415977.2020.1717488}

\bibitem{Cum-et-al-15}
P.~Cumsille, A.~Coronel, C.~Conca, C.~Qui{\~{n}}inao, C.~Escudero,
  \href{http://dx.doi.org/10.1186/s12976-015-0009-y}{Proposal of a hybrid
  approach for tumor progression and tumor-induced angiogenesis}, Theoretical
  Biology and Medical Modelling 12~(1) (2015) 13.
\newblock \href {https://doi.org/10.1186/s12976-015-0009-y}
  {\path{doi:10.1186/s12976-015-0009-y}}.
\newline\urlprefix\url{http://dx.doi.org/10.1186/s12976-015-0009-y}

\bibitem{China-WHO}
WHO, Report of the who-china joint mission on coronavirus disease 2019
  (covid-19),
  https://www.who.int/docs/default-source/coronaviruse/who-china-joint-mission-on-covid-19-final-report.pdf.

\bibitem{Hairer}
G.~W.~a. Ernst~Hairer, Syvert Paul~Nørsett,
  \href{http://gen.lib.rus.ec/book/index.php?md5=C0910AA6A5DA672512694EE0A8D1ED97}{Solving
  Ordinary Differential Equations I: Nonstiff Problems}, Springer Series in
  Computational Mathematics 8, Springer Berlin Heidelberg, 1987.
\newline\urlprefix\url{http://gen.lib.rus.ec/book/index.php?md5=C0910AA6A5DA672512694EE0A8D1ED97}

\bibitem{Maleewong2020}
M.~Maleewong,
  \href{http://medrxiv.org/content/early/2020/05/26/2020.05.23.20111500.abstract}{{Time
  delay epidemic model for COVID-19}}, medRxiv (2020) 2020.05.23.20111500\href
  {https://doi.org/10.1101/2020.05.23.20111500}
  {\path{doi:10.1101/2020.05.23.20111500}}.
\newline\urlprefix\url{http://medrxiv.org/content/early/2020/05/26/2020.05.23.20111500.abstract}

\bibitem{refId0}
{Elie, Romuald}, {Hubert, Emma}, {Turinici, Gabriel},
  \href{https://doi.org/10.1051/mmnp/2020022}{Contact rate epidemic control of
  covid-19: an equilibrium view}, Math. Model. Nat. Phenom. 15 (2020) 35.
\newblock \href {https://doi.org/10.1051/mmnp/2020022}
  {\path{doi:10.1051/mmnp/2020022}}.
\newline\urlprefix\url{https://doi.org/10.1051/mmnp/2020022}

\bibitem{data-nat}
M.~de~Ciencia Tecnología Conocimiento~e Innovación, Datos-covid19,
  https://github.com/MinCiencia/Datos-COVID19/blob/master/output/producto5/TotalesNacionales.csv
  (2020).

\bibitem{Baeza}
R.~Baeza-Yates, Work group of data sciences on covid-19 in chile,
  https://www.youtube.com/watch?v=T26vGwcbxH4\&feature=share\&fbclid= (2020).

\bibitem{data-reg}
M.~de~Ciencia Tecnología Conocimiento~e Innovación, Datos-covid19,
  https://github.com/MinCiencia/Datos-COVID19/blob/master/output/producto3/TotalesPorRegion.csv
  (2020).

\bibitem{Col-96}
T.~F. Coleman, Y.~Li, \href{https://doi.org/10.1137/0806023}{An interior trust
  region approach for nonlinear minimization subject to bounds}, SIAM Journal
  on Optimization 6~(2) (1996) 418--445.
\newblock \href {http://arxiv.org/abs/https://doi.org/10.1137/0806023}
  {\path{arXiv:https://doi.org/10.1137/0806023}}, \href
  {https://doi.org/10.1137/0806023} {\path{doi:10.1137/0806023}}.
\newline\urlprefix\url{https://doi.org/10.1137/0806023}

\bibitem{L.F-2001}
L.~S.~S. Thompson,
  \href{http://gen.lib.rus.ec/scimag/index.php?s=10.1016/s0168-9274(00)00055-6}{Solving
  ddes in matlab}, Applied Numerical Mathematics 37 (2001).
\newblock \href {https://doi.org/10.1016/s0168-9274(00)00055-6}
  {\path{doi:10.1016/s0168-9274(00)00055-6}}.
\newline\urlprefix\url{http://gen.lib.rus.ec/scimag/index.php?s=10.1016/s0168-9274(00)00055-6}

\bibitem{Banks}
H.~T. Banks, M.~Davidian, J.~R. Samuels, K.~L. Sutton,
  \href{https://doi.org/10.1007/978-90-481-2313-1_11}{An Inverse Problem
  Statistical Methodology Summary}, Springer Netherlands, Dordrecht, 2009, pp.
  249--302.
\newblock \href {https://doi.org/10.1007/978-90-481-2313-1_11}
  {\path{doi:10.1007/978-90-481-2313-1_11}}.
\newline\urlprefix\url{https://doi.org/10.1007/978-90-481-2313-1_11}

\bibitem{Greene}
W.~H. Greene, Econometric Analysis, seventh Edition, Pearson Education, 2012.

\bibitem{Banks-et-al08}
H.~Banks, S.~Dediu, S.~Ernstberger, {Sensitivity functions and their uses in
  inverse problems}, Journal of Inverse and Ill-posed Problems jiip 15~(7)
  (2008) 683--708.

\bibitem{Banks-et-al07}
H.~Banks, S.~Ernstberger, S.~Grove, {Standard errors and confidence intervals
  in inverse problems: sensitivity and associated pitfalls}, Journal of Inverse
  and Ill-posed Problems jiip 15~(1) (2007) 1--18.

\bibitem{rep}
S.~Núñez, A.~Meleán, They warn that chillán is going through the third peak
  of infections by covid-19 ({A}dvierten que {C}hillán atraviesa el tercer
  peak de contagios por {C}ovid-19),
  http://www.ladiscusion.cl/advierten-que-chillan-atraviesa-el-tercer-peak-de-contagios-por-covid-19/.

\bibitem{Ioan}
J.~Ioannidis, \href{https://www.who.int/bulletin/online_first/en/}{Infection
  fatality rate of covid-19 inferred from seroprevalence data}, Bulletin of the
  World Health Organization (2020).
\newblock \href
  {http://arxiv.org/abs/https://www.who.int/bulletin/online_first/BLT.20.265892.pdf?ua=1}
  {\path{arXiv:https://www.who.int/bulletin/online_first/BLT.20.265892.pdf?ua=1}}.
\newline\urlprefix\url{https://www.who.int/bulletin/online_first/en/}

\end{thebibliography}

\end{document}